\documentclass[aps,preprint,superscriptaddress,showpacs,preprintnumbers,amsmath,amssymb]{revtex4}
\usepackage{epsfig}

\usepackage{graphicx}               
\usepackage{dcolumn}                
\usepackage{multirow}               
\usepackage{amsmath}                
\usepackage{color}
\usepackage{comment}
\graphicspath{{ps}}                 
\renewcommand{\arraystretch}{1.1}   

\newcommand{\plcpipi}{{p {\overline\Lambda_c}^- \pi^+ \pi^-}}
\newcommand{\plcpi}{{p {\overline\Lambda_c}^- \pi^+}}
\newcommand{\plc}{{p {\overline\Lambda_c}^-}}
\newcommand{\plbar}{{p \overline{\Lambda}}}
\newcommand{\pipi}{\pi^+ \pi^-}
\newcommand{\plpipi}{{p \overline{\Lambda} \pi^+ \pi^-}}
\newcommand{\plpi}{{p \overline{\Lambda} \pi^-}}
\newcommand{\plrho}{{p \overline{\Lambda} \rho^0}}
\newcommand{\plf} {{p \overline{\Lambda} f_2(1270)}}

\newcommand{\mb}{{M_{\rm bc}}}
\newcommand{\de}{{\Delta{E}}}
\newcommand{\bp}{B^{+}}
\newcommand{\bz}{B^{0}}

\begin{document}
\preprint{\vbox{ \hbox{   }
                 \hbox{Belle Preprint 2009-21}
                 \hbox{KEK Preprint 2009-24}
}}

\title{ \quad\\[0.5cm] Observation of $\bp \to p \overline{\Lambda} \pi^+ \pi^-$ at Belle}

\affiliation{Budker Institute of Nuclear Physics, Novosibirsk}
\affiliation{Chiba University, Chiba}
\affiliation{University of Cincinnati, Cincinnati, Ohio 45221}
\affiliation{Department of Physics, Fu Jen Catholic University, Taipei}
\affiliation{The Graduate University for Advanced Studies, Hayama}
\affiliation{Hanyang University, Seoul}
\affiliation{University of Hawaii, Honolulu, Hawaii 96822}
\affiliation{High Energy Accelerator Research Organization (KEK), Tsukuba}
\affiliation{Institute of High Energy Physics, Chinese Academy of Sciences, Beijing}
\affiliation{Institute of High Energy Physics, Protvino}
\affiliation{Institute for Theoretical and Experimental Physics, Moscow}
\affiliation{J. Stefan Institute, Ljubljana}
\affiliation{Kanagawa University, Yokohama}
\affiliation{Korea University, Seoul}
\affiliation{Kyungpook National University, Taegu}
\affiliation{\'Ecole Polytechnique F\'ed\'erale de Lausanne (EPFL), Lausanne}
\affiliation{Faculty of Mathematics and Physics, University of Ljubljana, Ljubljana}
\affiliation{University of Maribor, Maribor}
\affiliation{Max-Planck-Institut f\"ur Physik, M\"unchen}
\affiliation{University of Melbourne, School of Physics, Victoria 3010}
\affiliation{Nagoya University, Nagoya}
\affiliation{Nara Women's University, Nara}
\affiliation{National Central University, Chung-li}
\affiliation{National United University, Miao Li}
\affiliation{Department of Physics, National Taiwan University, Taipei}
\affiliation{H. Niewodniczanski Institute of Nuclear Physics, Krakow}
\affiliation{Nippon Dental University, Niigata}
\affiliation{Niigata University, Niigata}
\affiliation{Novosibirsk State University, Novosibirsk}
\affiliation{Osaka City University, Osaka}
\affiliation{Panjab University, Chandigarh}
\affiliation{University of Science and Technology of China, Hefei}
\affiliation{Seoul National University, Seoul}
\affiliation{Sungkyunkwan University, Suwon}
\affiliation{School of Physics, University of Sydney, NSW 2006}
\affiliation{Excellence Cluster Universe, Technische Universit\"at M\"unchen, Garching}
\affiliation{Toho University, Funabashi}
\affiliation{Tohoku Gakuin University, Tagajo}
\affiliation{Department of Physics, University of Tokyo, Tokyo}
\affiliation{Tokyo Metropolitan University, Tokyo}
\affiliation{IPNAS, Virginia Polytechnic Institute and State University, Blacksburg, Virginia 24061}
\affiliation{Yonsei University, Seoul}

  \author{P.~Chen}\affiliation{Department of Physics, National Taiwan University, Taipei} 
  \author{M.-Z.~Wang}\affiliation{Department of Physics, National Taiwan University, Taipei} 
  \author{H.~Aihara}\affiliation{Department of Physics, University of Tokyo, Tokyo} 
  \author{A.~M.~Bakich}\affiliation{School of Physics, University of Sydney, NSW 2006} 
  \author{V.~Balagura}\affiliation{Institute for Theoretical and Experimental Physics, Moscow} 
  \author{V.~Bhardwaj}\affiliation{Panjab University, Chandigarh} 
  \author{M.~Bischofberger}\affiliation{Nara Women's University, Nara} 
  \author{A.~Bozek}\affiliation{H. Niewodniczanski Institute of Nuclear Physics, Krakow} 
  \author{M.~Bra\v cko}\affiliation{University of Maribor, Maribor}\affiliation{J. Stefan Institute, Ljubljana} 
  \author{T.~E.~Browder}\affiliation{University of Hawaii, Honolulu, Hawaii 96822} 
% ...

\collaboration{The Belle Collaboration} 

\noaffiliation

 \begin{abstract} 

We study charmless $B^+$ meson decays to the $p \overline{\Lambda} \pi^+ \pi^-$ final state using a $605\,{\rm fb}^{-1}$ data sample collected at the $\Upsilon(4S)$ resonance with the Belle detector at the KEKB asymmetric-energy $e^+ e^-$ collider. 
There are significant signals found with the $\plbar$ mass peaking near threshold.   
The observed branching fraction for non-resonant $\bp \to p \overline{\Lambda} \pi^+ \pi^-$  is $(5.92^{+0.88}_{-0.84} (stat.) \pm 0.69 (syst.)) \times 10^{-6}$ with a significance of 9.1 standard deviations. We also observe the intermediate three-body decay $\bp \to p \overline{\Lambda} \rho^0$ with a branching fraction of $(4.78^{+0.67}_{-0.64} (stat.) \pm 0.60 (syst.)) \times 10^{-6}$ and a significance of 9.5 standard deviations, and find a hint of a $\bp \to p \overline{\Lambda} f_2(1270)$ signal. No other intermediate three-body decay is found in this study.

 \end{abstract}  
\maketitle

\renewcommand{\thefootnote}{\fnsymbol{footnote}}
\setcounter{footnote}{0}
\clearpage

\section{Introduction}
The large data samples accumulated at the $B$ factories allows us to study the properties of baryonic $B$ meson decays, which are less well understood than $B$ decays into mesons.
An interesting hierarchy has been established experimentally in branching fractions of baryonic $B$ decays, namely
$\mathcal{B}(\bz \to \plcpipi) > \mathcal{B}(\bp \to \plcpi) > \mathcal{B}(\bz \to \plc)$~\cite{CLEO,Gaby1,Gaby2,Park}.
The above decays proceed via $ b \to c$ tree diagrams.
The corresponding hierarchy has not yet been confirmed in charmless baryonic $B$ decays, which presumably proceed via $b \to s$ penguin or $b \to u$ tree diagrams. 
So far, many three-body charmless baryonic  $B$ decays have been observed but only very stringent upper limits have been set for quasi-two-body decays~\cite{PDG}. 
One intriguing  experimental finding in charmless baryonic three-body decays is that all the baryon-antibaryon mass distributions peak near threshold.
Many theoretical investigations, using QCD counting rules~\cite{HandC,gengg} or pole models~\cite{Rho,ChengYang} 
in the factorization approach, have been carried out for charmless baryonic $B$ decays with $\plbar$ in the final state. 
These theoretical results on the branching fractions and the shape of the baryon-antibaryon mass spectrum agree reasonably well with the experimental findings. However, the proton angular distribution of the threshold peak observed in $\bz \to \plpi$~\cite{Wang} seems to violate the short-distance $b \to s$ picture and awaits a theoretical explanation. Since the $\Lambda$ hyperon could be a useful tool to probe the $b \to s$ process~\cite{HandC,Suzuki} and the related three-body decays can be used for a T-violation study~\cite{TV}, it is of general interest to search for
additional charmless baryonic decay modes.

In this paper, we present for the first time a study of the four-body charmless baryonic decay $\bp \to \plpipi$~\cite{conjugate} in order to  test the multi-body hierarchy in the charmless case. 
To investigate the threshold enhancement effect found in three-body decays, we study partial branching fractions as a function of the baryon-antibaryon mass.
We also search for possible intermediate three-body decays to the same $\plpipi$ final states, such as $\plrho$.

\section{Event Selection and Reconstruction}
\subsection{Data Samples and the Belle Detector}
The data sample used in this study corresponds to an integrated luminosity of 605 fb$^{-1}$ and contains 657 $ \times 10^6~B\overline{B}$ pairs collected with the Belle detector on the $\Upsilon(4S)$ resonance at the KEKB asymmetric-energy $e^+e^-$ (3.5~GeV  and 8~GeV) collider~\cite{KEKB}.

The Belle detector is a large-solid-angle magnetic spectrometer that consists of a silicon vertex detector (SVD),
a 50-layer central drift chamber (CDC), an array of aerogel threshold Cherenkov counters (ACC), a barrel-like arrangement of time-of-flight scintillation counters (TOF), and an electromagnetic calorimeter composed of CsI(Tl) crystals located inside a superconducting solenoid coil that provides a 1.5~T magnetic field. 
An iron flux-return located outside of the coil is instrumented to detect $K_L^0$ mesons and to identify muons. 
The Belle detector is described in detail elsewhere~\cite{Belle}.

\subsection{Selection Criteria and $B$ Meson Reconstruction}
The selection criteria for the final state particles of $\bp \to \plpipi$ are based on information obtained from the tracking system (SVD and CDC) and the hadron identification system (CDC, ACC, and TOF). 
We use the same requirements as those described in Ref.~\cite{Wang}. Candidate $B$ mesons are identified with the following two kinematic variables in the center-of-mass (CM) frame: the beam-energy-constrained mass $\mb = \sqrt{E^2_{\rm beam}-p^2_B}$, and the energy difference $\de = E_B - E_{\rm beam}$, where $E_{\rm beam}$ is the beam energy, and $p_B$ and $E_B$ are the momentum and energy, respectively, of the reconstructed $B$ meson.
We define a candidate region as 5.20 GeV/$c^2 < \mb < 5.3$ GeV/$c^2$ and $-0.1$ GeV $ < \de< 0.5$ GeV. 
The lower bound in $\de$ is chosen to exclude all backgrounds due to baryonic $B$ decays with higher multiplicities.
From a GEANT~\cite{geant} Monte Carlo (MC) simulation, the signal peaks in a signal box defined by the requirements 
5.27 GeV/$c^2 < \mb < 5.29$ GeV/$c^2$ and $|\de|< 0.05$ GeV.

\subsection{Background Suppression}
After the above selection requirements, the background in the candidate region arises predominantly from continuum $e^+e^- \to q\overline{q}$ ($q = u,\ d,\ s,\ c$) processes.
We suppress jet-like continuum background relative to the more spherical $B\overline{B}$ signal using a Fisher discriminant~\cite{fisher} that combines seventeen event shape variables as described in Ref.~\cite{ksfw}.
We then optimize the coefficients separately in seven different missing-mass regions defined in Ref.~\cite{chang}
to improve the signal-to-background ratio.
Probability density functions (PDFs) for the Fisher discriminant and the cosine of the angle between the $B$ flight direction and the beam direction in the $\Upsilon(4S)$ rest frame are combined to form the signal (background)
likelihood ${\mathcal L}_{s}$ (${\mathcal L}_{b}$).
The signal PDFs are determined using signal MC simulation; 
the background PDFs are obtained from the sideband region of the data:
$5.20$ GeV/$c^2$  $ < \mb < 5.26$ GeV/$c^2$ or $\left| \de \right| > 0.1$ GeV.
We require the likelihood ratio ${\mathcal R} = {\mathcal L}_s/({\mathcal L}_s+{\mathcal L}_b)$ to be greater than 0.85 for all seven missing-mass regions. 
This value is determined by optimizing $n_s/\sqrt{n_s+n_b}$ as a function of ${\mathcal R}$, where $n_s$ and $n_b$ denote the expected numbers of signal and background events in the signal box, respectively. 
We assume a signal branching fraction of $ 10^{-5}$ to estimate $n_s$ and use the data sideband events to determine $n_b$. 

To ensure that the decay process is genuinely charmless, we apply a charm veto. 
$\bp \to \plpipi$ candidate events with $2.10$ GeV/$c^2 < M_{\Lambda\pi^+} < 2.32$ GeV/$c^2$ are removed to avoid background from $B^+ \to p {\overline\Lambda_c}^- \pi^+$ and $B^+ \to p {\overline\Sigma_c}^0$, ${\overline\Sigma_c}^0 \to {\overline\Lambda_c}^- \pi^+$ with ${\overline\Lambda_c}^- \to \overline\Lambda \pi^-$ decay.                     
From MC simulation, there are events from $B^0 \to p \overline{{\Lambda}} \pi^-$ in the candidate region.
For simplicity, we remove $B$ candidates if the corresponding reconstructed $\mb$ and $\de$, using only
the three daughters (i.e., $p\overline{\Lambda}\pi^- $), are in the signal box. This selection removed about 4.8\% of candidate events.  The contribution of the $B$ background
component with $\Sigma \to \Lambda\gamma$ has a $\de$ distribution that is different from signal..
This is  included in the systematic uncertainty from PDF modeling by comparing the fit results with and 
without this background component in the fit. If there are  multiple $B$ candidates in a single event, 
we select the one with the best ${\chi^2}$ value for the $p\pipi$ vertex fit.
The fraction of multiple $B$ candidate events is 18.8\%.
The systematic errors due to multiple $B$ candidates are described later.

\section{Extraction of Signal}
In order to obtain the signal yield for the $\plpipi$ final state, we perform an unbinned extended likelihood fit 
that maximizes the likelihood function 
\begin{eqnarray}
\nonumber L=&&\frac{e^{-(N_{p\overline\Lambda\pi^+\pi^-}+N_{q\overline{q}})}}{N!} \\
\nonumber   &&\times \prod_{i=1}^{N} (N_{p\overline\Lambda\pi^+\pi^-} P_{p\overline\Lambda\pi^+\pi^-}
+N_{q\overline{q}} P_{q\overline{q}}),
\end{eqnarray}
where $N$ is the number of total events. $N_{\plpipi}$ and $N_{q\overline{q}}$
are fit parameters representing the numbers of $\plpipi$ signal events and continuum background
events, respectively. 
Each function $P$ is a PDF expressed as uncorrelated products of shapes of the $\mb$ and $\de$ distributions: $P =P_{\mb} \times P_{\de}$.

For the PDFs related to $B$ decays, 
we use a Gaussian function to represent $P_{\mb}$ and a double Gaussian for $P_{\de}$ 
with parameters determined from a MC signal simulation of $\bp \to \plpipi$ phase space.  
To model the $P_{\mb}$ continuum background, we use a parameterization that was first employed by the ARGUS collaboration, 
$ f(\mb)\propto x\sqrt{1-x^2}e^{-\xi (1-x^2)}$,   
where $x$ is $\mb/E_{\rm beam}$ and $\xi$ is a fit parameter~\cite{argus}. 
The $P_{\de}$ continuum background shape is modeled by a normalized second-order polynomial 
whose coefficients are fit parameters.

\begin{figure}[htb]
 \begin{center}
{\includegraphics[width=0.45\textwidth]{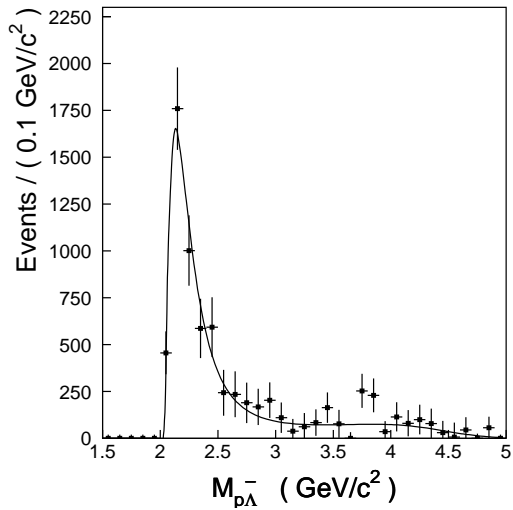}}
 \caption{
 $B$ signal yields obtained as a function of the $\plbar$ mass. The yields are
corrected for the mass-dependent efficiency. The solid curve shows a fit with a threshold function.
}
 \label{fg:m12_eff}
  \end{center}
\end{figure}

We study mass spectra for the $\plbar$, $p \pi^-$, $p \pi^+$,
$\Lambda\pi^-$, $\Lambda\pi^+$, $\pipi$, $\plbar\pi^-$, $\plbar\pi^+$, $p\pipi$ and $\Lambda\pipi$
combinations using the $B$ signal yields obtained as functions of those masses.
There is a clear enhancement near threshold in the $\plbar$ mass for signal candidates. 
We fit the $\plbar$ mass distribution with a threshold function: 
$f_{\textrm{thr}} \propto (\Delta{M_\plbar})^{s}\times
e^{[c_1\times(\Delta{M_\plbar})+c_2\times(\Delta{M_\plbar})^2+c_3\times(\Delta{M_\plbar})^3]}$, where $\Delta{M_\plbar}\equiv M_{\plbar}-m_{p}-m_{\Lambda} $; $~s,~c_1,~c_2$ and $c_3$ are fit parameters.
The result is shown in Fig.~\ref{fg:m12_eff}. The excess in the region $3.5$ GeV/$c^2 < M_{\plbar} < 4.0$ GeV/$c^2$ will be 
investigated in a study of systematic effects.
The only other resonance-like structure found is in the $\pipi$ mass spectrum where
we observe a clear $\rho^0$ signal. We correct the fitted $B$ signal yields by the efficiencies to obtain the total ${\mathcal B}(\bp \to \plpipi)_{\rm Tot} = (11.28^{+0.91}_{-0.72} \pm 1.03) \times 10^{-6}$, where non-resonant and possible intermediate resonance decays are all included.

In order to investigate three-body decays such as $\bp \to \plrho$ and $\bp \to \plf$, we perform an unbinned extended 
likelihood fit that maximizes the likelihood function 
\begin{eqnarray}
\nonumber L &&=\frac{e^{-(N_{p\overline\Lambda\pi^+\pi^-}+N_{p\overline\Lambda\rho}+N_{p\overline\Lambda f_2}+N_{q\overline{q}})}}{N!} \\ \nonumber    &&\times \prod_{i=1}^{N}(N_{p\overline\Lambda\pi^+\pi^-}  P_{p\overline\Lambda\pi^+\pi^-}
               +N_{p\overline\Lambda\rho}P_{p\overline\Lambda\rho}\\
\nonumber    &&+N_{p\overline\Lambda f_2} P_{p\overline\Lambda f_2} +N_{q\overline{q}} P_{q\overline{q}}),
\end{eqnarray}
where $N$ is number of total events. Here $N_{\plpipi}$, $N_{\plrho}$, $N_{\plf}$, and $N_{q\overline{q}}$
are fit parameters representing the yields of non-resonant $\plpipi$, $\plrho$ and $\plf$ signal and continuum background contributions , respectively. 
Each function $P$ is a PDF expressed as uncorrelated products of shapes of the $\mb$, $\de$ and $M_{\pipi}$ distributions: $P =P_{\mb} \times P_{\de} \times P_{M_{\pipi}} $.

For the PDFs related to $B$ decays,
we use a Gaussian function to represent $P_{\mb}$ and a double Gaussian for $P_{\de}$ 
with parameters determined from MC signal simulation. For the continuum background PDF, 
we use an ARGUS function to model the $P_{\mb}$ distribution while the $P_{\de}$ distribution is modeled by a normalized second-order polynomial whose coefficients are fit parameters.

The PDFs for the $M_{\pipi}$ mass distributions in $\bp \to \plrho$ and $\bp \to \plf$ are 
smoothed histograms obtained from MC simulation using a sample with the $\plbar$ 
shape fixed from data and  Breit-Wigner forms for the $\rho^0$ or $f_2$ resonances. 
The subsequent decay angular distribution is assumed to be flat.   
The PDF $P_{M_{\pipi}}$ for $\bp \to \plpipi$ 
is also a smoothed histogram from a three-body
phase space MC sample, which again follows a threshold 
shape in the $\plbar$ mass spectrum for data. 
The continuum background PDF for $M_{\pipi}$
is modeled by the sum of smoothed histogram functions taken from Breit-Wigner distributions for $\rho^0$ 
and $f_0$ resonances, and a threshold function: 
$P_{M_{\pipi}} = r_1\times P_{\rho} + r_1\times P_{f_0}
+ (1-r_1- r_2)\times P_{\textrm{thr}}$ and
$P_{\textrm{thr}} \propto (\Delta{M_{\pipi}})^{s}\times
e^{[c_1\times(\Delta{M_{\pipi}})+c_2\times(\Delta{M_{\pipi}})^2]}$
where $\Delta{M_{\pipi}}\equiv M_{\pipi}-2m_{\pi}$; $r_1,~r_2,~s,~c_1$ and $c_2$ are fit parameters.

\renewcommand{\arraystretch}{1.2} 

\begin{table}
\caption{Systematic errors (in \%).}
\begin{center}
\begin{tabular}{l|ccc}
\hline
\hline
~~~Source                    &      $~~~\plpipi$ & ~~$\plrho$~~ & ~~$\plf$~~ \\
\hline
Tracking                          & 6.0         & 5.7        & 5.7\\
Proton ID                         & 4.2         & 4.2        & 4.2\\
Charged Pion ID                   & 1.3         & 1.7        & 1.4\\
$\Lambda$ reconstruction          & 3.3         & 2.8        & 3.0\\
${\mathcal B}(\Lambda \to p\pi^+) $        & 0.8         & 0.8        & 0.8\\
${\mathcal B}(\rho/f_2 \to \pipi) $        & --          & 0.2        & 2.8\\
$\mathcal R$ Selection            & 2.4         & 2.4        & 2.4\\
Multiple Counting                 & 3.0         & 3.0        & 3.0\\
PDF Modeling                      & 7.3         & 9.0        & 9.6\\
Number of $B\overline{B}$ Pairs        & 1.4         & 1.4        & 1.4\\
\hline
Total                             & 11.7       & 12.6       & 13.4\\ 
\hline
\hline
\end{tabular}
\end{center}
\label{systematicxx}
\end{table}

\section{Results}
\begin{figure*}[htb]
 \vskip 0.10cm
 \begin{center}
 \hskip -6.5cm {\bf (a)} \hskip 6.7cm {\bf (b)}\hskip 5.5cm\\
 \vskip -0.6cm
 \includegraphics[width=0.4\textwidth]{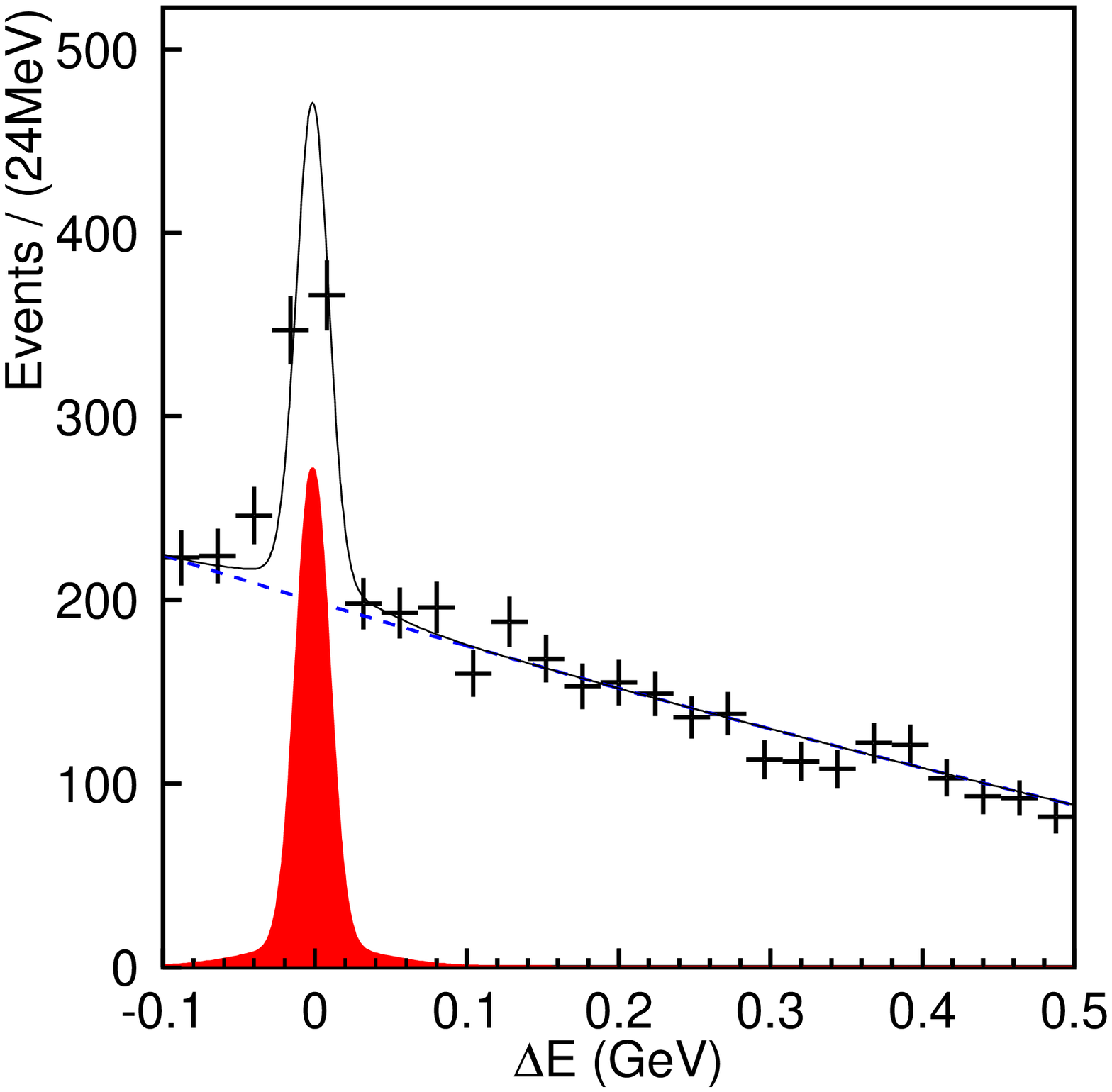}
 \includegraphics[width=0.4\textwidth]{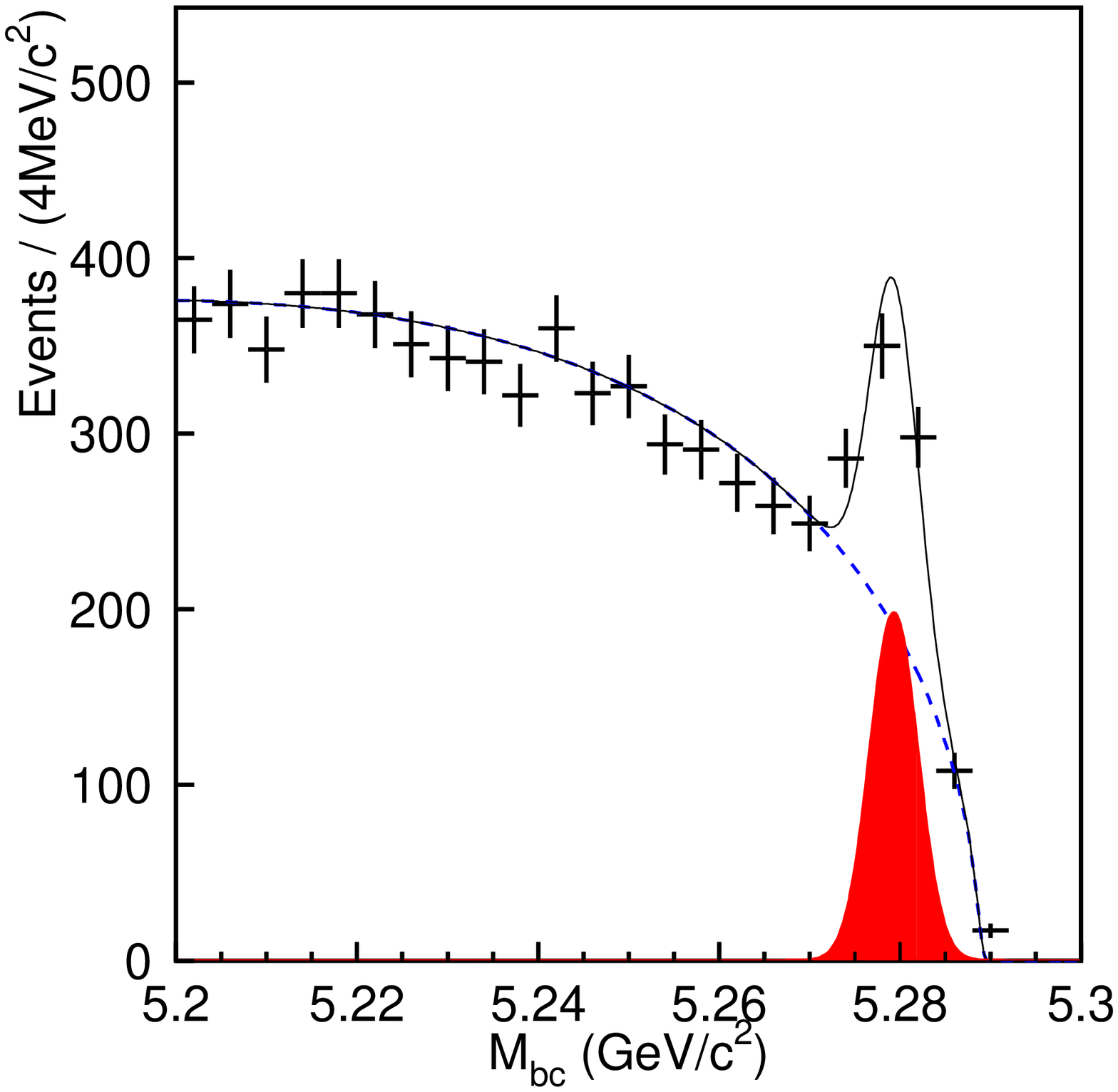}\\
 \hskip -6.6 cm{\bf{(c)}}
 \vskip -0.6cm
 \includegraphics[width=0.4\textwidth]{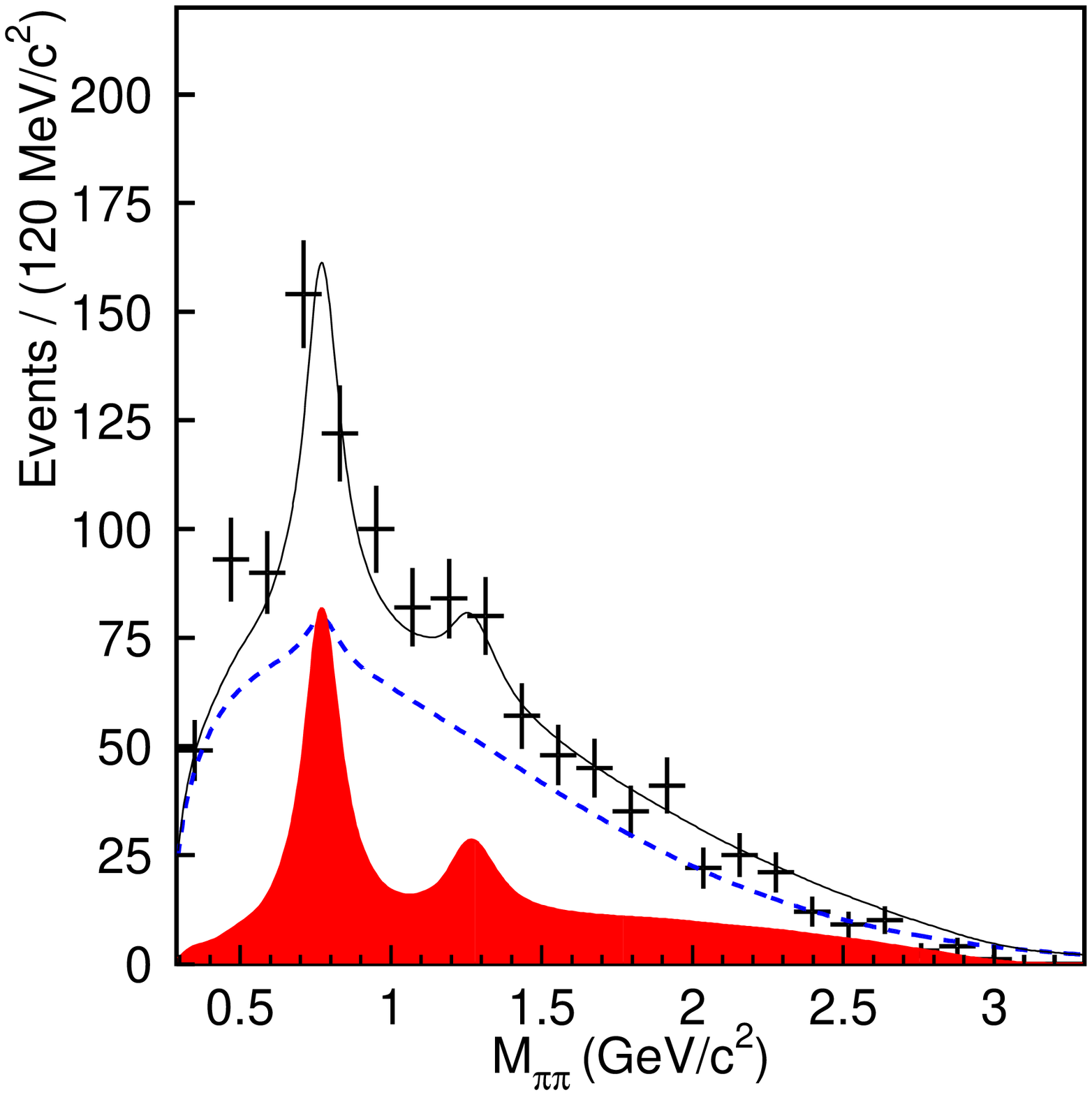}
  \caption{Distributions of
 (a) $\de$ (with $5.27$ GeV/$c^2 < \mb < 5.29$ GeV/$c^2$),
 (b) $\mb$ (with $|\de| < 0.05$ GeV) and (c) $M_{\pipi}$ (with $5.27$ GeV/$c^2 < \mb < 5.29$ GeV/$c^2$
and $|\de| < 0.05$ GeV).  The solid curve represents the fit projection,
which is the sum of signal and background (dashed curve) estimates.
The shaded area represents the sum of signal components i.e. $\bp \to \plpipi$, $\bp \to \plrho$ and $\bp \to \plf$.
 }
 \label{fg:mergembde-1}
  \end{center}
\end{figure*}

Figure ~\ref{fg:mergembde-1} shows the fit results for $\bp \to \plpipi$.
The resulting signal yields are $167.8^{+25.0}_{-23.7}
$, $131.2^{+18.3}_{-17.5}$ and $39.1^{+14.9}_{-14.0}$ with statistical significances of 9.1, 9.5, and 
3.0 for $\bp \to \plpipi$, $\bp \to \plrho$ and $\bp \to \plf$, respectively.
The significance is defined as $\sqrt{-2{\rm ln}(L_0/L_{\rm max})}$,
where $L_0$ and $L_{\rm max}$ are the likelihood values returned by the 
fit with the signal yield fixed to zero and at its best fit value, respectively.

The branching fractions are calculated by using the formula
\begin{eqnarray}
\nonumber \mathcal{B} =\frac{N_{signal}}{\epsilon \times N_{B\overline{B}}  },
\end{eqnarray}
where $N_{signal}$, $\epsilon$ and $N_{B\overline{B}}$ are the number of signal events, the efficiency estimated from MC simulation and the number of $B\overline{B}$ pairs, respectively.
We obtain
${\mathcal B}(\bp \to \plpipi) = (5.92^{+0.88}_{-0.84} \pm 0.69) \times 10^{-6}$, 
${\mathcal B}(\bp \to \plrho ) = (4.78^{+0.67}_{-0.64} \pm 0.60) \times 10^{-6}$, and 
${\mathcal B}(\bp \to \plf   ) = (2.03^{+0.77}_{-0.72} \pm 0.27) \times 10^{-6}$  
with efficiencies of 4.32\%, 4.17\% and 2.94\%, where sub-decay branching fractions are included. 
To verify the validity of the branching fractions, 
we select $\bp \to \plcpi$ events with $2.10$ GeV/$c^2 < M_{\Lambda\pi^+} < 2.32 $ GeV/$c^2$ for
$\Lambda_c^+$ and use a 2D ($\mb - \de$) extended likelihood fit method.  
We find the branching fraction of $\bp \to \plcpi$ to be
$(2.44 \pm 0.30 (stat.)) \times 10^{-6}$,
which agrees with the value calculated using world averages~\cite{PDG}, $\mathcal{B}(\bp \to \plcpi , \Lambda_c^+ \to \Lambda \pi^+)= (2.25 \pm 0.87) \times 10^{-6}$.

\renewcommand{\arraystretch}{1.4} 
\addtolength{\tabcolsep}{5pt}
\begin{table*}
\caption{Summary.}
\begin{center}
\begin{tabular}{c|rccc}
\hline
\hline
~~~~~Mode~~~~~                      &~~~~Yield~~~~                     & ~~~~~Efficiency(\%) &  ${\mathcal B (10^{-6})}$ & ~~~~~Significance\\
\hline
~~~~~$\plpipi$~~~~~            & ~~~~$167.8^{+25.0}_{-23.7}$  &  4.32          & $~~~~~5.92^{+0.88}_{-0.84} \pm 0.69 $               & ~~~~~9.1\\
~~~~~$\plrho$~~~~~                  & ~~~~$131.2^{+18.3}_{-17.5}$  &  4.17          & $~~~~~4.78^{+0.67}_{-0.64} \pm 0.60 $       & ~~~~~9.5\\
~~~~~$\plf$~~~~~                    & ~~~~$ 39.1^{+14.9}_{-14.0}$  &  2.94          & $~~~~~2.03^{+0.77}_{-0.72} \pm 0.27 $       & ~~~~~3.0\\
\hline
\hline
\end{tabular}
\end{center}
\label{summaryxx}
\end{table*}

\renewcommand{\arraystretch}{1.0}

\section{Systematic uncertainties}
Systematic uncertainties 
are estimated using high-statistics control data samples.
The tracking efficiencies are measured with
fully and partially reconstructed $D^*$ samples.
For proton identification, we use a  $\Lambda \to p \pi^-$ sample, while for
$K/\pi$ identification we use a $D^{*+} \to D^0\pi^+$,
 $D^0 \to K^-\pi^+$ sample. The average
efficiency difference for  particle identification  (PID)
between data and MC has been corrected to obtain the final branching fraction
measurements. 
The corrections are 14.8\%, 13.8\% and 14.2\% for $\bp \to \plpipi$, $\bp \to \plrho$ and $\bp \to \plf$, respectively.
The uncertainties associated with
the PID corrections are estimated to be 4.2\% for two protons (one from $\Lambda$ decay)
and 1--2\% for two charged pions.

For $\Lambda$ reconstruction, we have additional
uncertainties of $3.0\%$, $2.4\%$ and $2.6\%$ for
$\plpipi$, $\plrho$ and $\plf$, for the
efficiencies of tracks displaced from the interaction point. These uncertainties are
determined from the
difference between $\Lambda$ proper time distributions for data and MC
simulation. 
There are also uncertainties of $1.2\%$, $1.3\%$ and $1.3\%$ 
due to the $\Lambda$ mass selection and $0.7\%$, $0.5\%$ and $0.6\%$ due to the
$\Lambda$ vertex selection, for the $\plpipi$, $\plrho$ and $\plf$ modes, respectively. 

The uncertainty in the likelihood ratio  requirement, $\mathcal R$, is estimated from
a control sample ($\bp \to \overline{D}^0 \pi^+$, $\overline{D}^0 \to K^+ \pi^- \pi^- \pi^+$), 
which has the same number of final state particles. 
For the multiple counting uncertainty, we take the difference of branching
fractions with and without the best candidate selection as the systematic error. 

Fitting uncertainty is checked by considering different PDF models. 
MC statistics contribute a 1\% systematic error.
According to our MC simulation study, the charmless $B$ decay that most affects our  signal determination is $\bp \to p \overline{\Sigma} \pi^+ \pi^- $. This mode contributes 1.5\%, 0.5\% and 0.5\% uncertainties for $\bp \to \plpipi$, $\bp \to \plrho$ and $\bp \to \plf$, respectively.
In the $M_{\pipi}$ spectrum, we use a threshold function and a phase space shape to model the generic $\bp \to \plpipi$ process.
The modeling of the threshold function shape in the $M_{\pipi}$ spectrum contributes systematic errors of
6.9\%, 8.8\% and 9.0\% for $\bp \to \plpipi$, $\bp \to \plrho$ and $\bp \to \plf$.
Another uncertainty is due to MC modeling of the $M_{\plbar}$ distribution and  is included by varying the fit parameter of the threshold function and adding a phase space component in the fit.  
The uncertainties due to $M_{\plbar}$ modeling are 1.3\%, 1.5\% and 3.2\% for $\bp \to \plpipi$, $\bp \to \plrho$ and $\bp \to \plf$, respectively. We neglect the uncertainty due to interference effects between resonances and $\pipi$, assuming the non-resonant $\pipi$ system is predominately in an $S$-wave.
 
We assume  that the branching fractions of $\Upsilon(4S)$ to neutral and charged $B\overline{B}$ pairs are equal.
The correlated errors are added linearly and the uncorrelated ones are added in quadrature. 
The total systematic uncertainties are 11.7\%, 12.6\% and 13.4\% for
the $\plpipi$, $\plrho$ and $\plf$ modes, respectively, and 
the details of each decay channel are summarized in Table~\ref{systematicxx}.

\section{Summary}
Using 657 $ \times 10^6 ~ B\overline{B}$ events, we 
observe non-resonant $\bp \to \plpipi$ and $\bp \to \plrho$ decays with significances of 9.1 and 9.5 standard deviations as shown in Table~\ref{summaryxx}.
A hint of a $\bp \to \plf$ signal is found. 
This is the first observation of 
a  four-body charmless baryonic $B$ decay.   
We also observe low-mass $\plbar$ threshold enhancements in both the $\plpipi$ and $\plrho$ modes. 
The branching fraction of the four-body decay is comparable to the corresponding three-body
decays modes; however, the central values may indicate that the hierarchy 
established in the charmed case still holds  for charmless baryonic decay.
The observed branching fraction for $\bp \to \plrho$ 
is comparable to that for $\bz \to \plpi$, and is about an 
order of magnitude larger than the theoretical prediction~\cite{Rho}.

We thank the KEKB group for the excellent operation of the
accelerator, the KEK cryogenics group for the efficient
operation of the solenoid, and the KEK computer group and
the National Institute of Informatics for valuable computing
and SINET3 network support.  We acknowledge support from
the Ministry of Education, Culture, Sports, Science, and
Technology (MEXT) of Japan, the Japan Society for the 
Promotion of Science (JSPS), and the Tau-Lepton Physics 
Research Center of Nagoya University; 
the Australian Research Council and the Australian 
Department of Industry, Innovation, Science and Research;
the National Natural Science Foundation of China under
contract No.~10575109, 10775142, 10875115 and 10825524; 
the Department of Science and Technology of India; 
the BK21 program of the Ministry of Education of Korea, 
the CHEP src program and Basic Research program (grant 
No. R01-2008-000-10477-0) of the 
Korea Science and Engineering Foundation;
the Polish Ministry of Science and Higher Education;
the Ministry of Education and Science of the Russian
Federation and the Russian Federal Agency for Atomic Energy;
the Slovenian Research Agency;  the Swiss
National Science Foundation; the National Science Council
and the Ministry of Education of Taiwan; and the U.S.\
Department of Energy.
This work is supported by a Grant-in-Aid from MEXT for 
Science Research in a Priority Area ("New Development of 
Flavor Physics"), and from JSPS for Creative Scientific 
Research ("Evolution of Tau-lepton Physics").

\clearpage


\begin{thebibliography}{99}
\bibitem{conjugate} {Throughout this report, inclusion of 
charge-conjugate modes is always implied unless otherwise stated.}
\end{thebibliography}
\end{document}